\documentclass[twocolumn,secnumarabic,amssymb, nobibnotes, aps, superscriptaddress, prb, reprint]{revtex4-1}
\usepackage{graphicx}% Include figure files
\usepackage{bm}% bold math
\usepackage{float}% float fgures
\usepackage{braket}
\usepackage{upgreek}

\begin{document}

\title{Decoherence measurements in crystals of molecular magnets}

\author{Gheorghe Taran}%
\email[]{gheorghe.taran@kit.edu}
\affiliation{Physikalisches Institute, KIT, Wolfgang-Gaede-Str. 1, Karlsruhe D-76131}
%\affiliation{N\'eel Institute, CNRS, 25 rue des Martyrs,Grenoble 38042}
\author{Edgar Bonet}%
\email[]{bonet@grenoble.cnrs.fr}
\affiliation{Univ. Grenoble Alpes, CNRS, Grenoble INP\footnote{Institute of Engineering Univ. Grenoble Alpes}, Institut N\'eel, 38000 Grenoble, France}
\author{Wolfgang Wernsdorfer}%
\email[]{wolfgang.wernsdorfer@kit.edu}
\affiliation{Physikalisches Institute, KIT, Wolfgang-Gaede-Str. 1, Karlsruhe D-76131}
\affiliation{Univ. Grenoble Alpes, CNRS, Grenoble INP\footnote{Institute of Engineering Univ. Grenoble Alpes}, Institut N\'eel, 38000 Grenoble, France}
\affiliation{Institute of Nanotechnology (INT), Karlsruhe Institute of Technology (KIT), Hermann-von-Helmholtz-Platz 1, D-76344 Eggenstein-Leopoldshafen}

\date{\today}%

\begin{abstract}
Decoherence processes in crystals of molecular magnets are prototypical for interacting electronic spin systems. We analyze the Landau-Zener dynamics of the archetypical TbPc$_2$ complex diluted in a diamagnetic monocrystal.
The dependence of the tunneling probability on the field sweep rate is evaluated in the framework of the recently proposed master equation in which the decoherence processes are described through a phenomenological Lindblad operator. Thus, we showcase low temperature magnetic measurements that complement resonant techniques in determining small tunnel splittings and dephasing times.
\end{abstract}

\maketitle

\textit{Introduction.---} 
The coherent dynamics of an ensemble of weakly coupled spin systems is central to both the development of mesoscopic quantum physics~\cite{stamp2012environmental,takahashi2011decoherence} and to the fast advancing field of quantum engineering~\cite{stamp2009,troiani2011molecular,ghirri2017molecular}.
Amongst solid state electron spin systems that are researched as potential quantum bits, (\textit{e.g.} semiconductor quantum dots~\cite{hanson2007spins}, nitrogen vacancies centers in diamond~\cite{awschalom2007diamond}, molecular magnets, phosphorus or bismuth in silicon~\cite{morton2011embracing}), molecular magnets proved to be especially useful model systems.
Thus, many purely quantum phenomena like ground state tunneling~\cite{friedman1996macroscopic}, phonon and photon assisted tunneling transitions~\cite{thomas1996macroscopic,wernsdorfer2004resonant}, spin parity and quantum phase interference~\cite{wernsdorfer1999quantum}, phase coherence and Rabi oscillations~\cite{bertaina2008quantum} were analyzed in great detail in these systems.
When it comes to the study of decoherence, the common ground between different qubit systems is found in the description of the environment by standardized models like oscillatory or spin baths~\cite{stamp2012environmental}.
The main advantages of molecular magnets arise from their diversity and chemical tunability of the spin ground state and the intra- and intermolecular interactions (\textit{e.g.} through the appropriate choice of the organic ligands)~\cite{liu2018symmetry,mcadams2017molecular}.
Thus, one of the best characterized molecular system, the Fe$_8$ complex~\cite{friedman2010single}, was used to validate the theory of environmental decoherence  against experiment~\cite{takahashi2011decoherence}. 
In the above experiment as well as the breakthrough achievements like the first measurements of the spin relaxation times~\cite{ardavan2007will} and 
the observation of millisecond coherence time and Rabi
oscillations at room temperature~\cite{zadrozny2015millisecond, atzori2016room, tesi2016quantum}, electron paramagnetic resonance (EPR) was the technique of \textit{choice}. 
However, the stringent requirement for a system to be susceptible to EPR investigation is a large coherence time~\cite{takahashi2009coherent}. 
Through this work we show that using incoherent Landau-Zener tunneling dynamics~\cite{troiani2017landau} we are able to determine the intrinsic tunneling time and the decoherence rate, thus complementing the resonance techniques and providing a new tool to probe the quantum properties of molecules that could be candidates for implementing quantum bits.

We analyze the magnetic response characterizing a diluted crystal of terbium(III) bis(phtalocyanine) (TbPc$_2$) lanthanide single ion molecular magnets (SIMMs).
Amongst already numerous example of SIMMs, the TbPc$_2$ complex makes its claim to fame through
events  central to the development of both
the field of single molecule magnets and molecular spintronics.
First, the analysis of its magnetic bistability proved that
a molecular complex with a single magnetic center can exhibit
a large effective energy barrier~\cite{ishikawa2003lanthanide}.
Second, the possibility to dilute the TbPc$_2$ molecules in
a isostructural diamagnetic matrix and
the strong hyperfine interaction, characteristic of 
lanthanide ions, allowed to experimentally evidence 
the resonant quantum tunneling between
mixed states of electronic and nuclear origin~\cite{ishikawa2005quantum}.
Furthermore, the planar structure of the molecule made possible 
its deposition on different substrates, and
thus subsequent inclusion in spintronics devices~\cite{cornia2003direct,komeda2011observation}.
The last point is especially relevant in the race to develop
quantum information processing devices~\cite{troiani2011molecular}.
An example of the symbiotic relationship between 
the fundamental research and technological application is given by 
the TbPc$_2$ single molecule spin transistor.
It was first used to read out and control
both the electronic and the nuclear spin~\cite{thiele2014electrically},
then to successfully implement quantum algorithms~\cite{godfrin2017operating}.
The same device provided the experimental means to 
explore how the effective character of the resonant tunneling changes when
the dephasing of environmental or measurement origin is taken into account~\cite{troiani2017landau}.

We start by reviewing the low temperature magnetic properties of the TbPc$_2$ complex and the Landau-Zener formalism in which it's dynamics is studied.
Then, we describe the novel method by which we obtain the dependence of the tunneling transitions on the sweeping rate, alongside presenting the first experimental evidence of the thermalization of the $^{159}$Tb nuclear spins. 
Finally, the phenomenological model proposed in Ref.~\onlinecite{troiani2017landau} is used to motivate the dynamics of the  molecular spin and to study it quantitatively leading to experimental estimates of both the intrinsic tunneling time and the dephasing time.

\begin{figure}
    \includegraphics[width=0.48\textwidth]{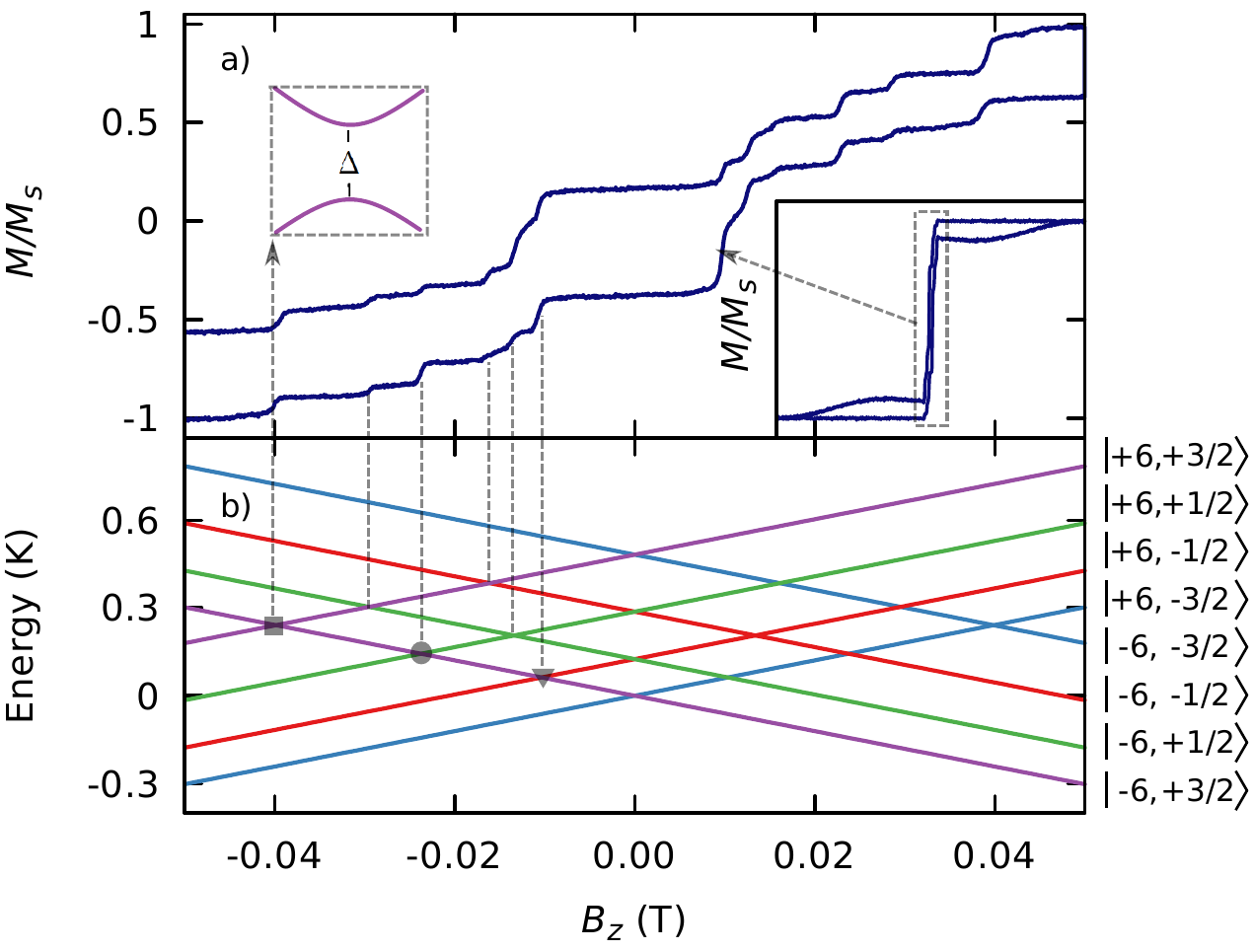}
    \caption{
    \label{fig:zd}
    \text{a)} A zoom on the magnetic hysteresis loop of a crystal containing TbPc$_2$ SIMMs diluted in a diamagnetic, isostructural matrix formed by YPc$_2$ molecules, with [TbPc$_2$]/[YPc$_2$] ratio of 1\%, measured by microSQUID technique~\cite{wernsdorfer2009}. The insets show: (\textit{left}) a zoom of a level anticrossing between two hyperfine states and (\textit{right}) the entire hysteresis loop.
    \text{b)} Hyperfine structure of the lowest doublet, $m_J=\pm6$, as a function of an applied magnetic field.
	}
	\vspace{-15pt}
\end{figure}

\textit{Theory.---} 
The magnetic properties of the TbPc$_2$ molecule are dominated by the coupling between the spin ($S=3$) and orbital ($L=3$) angular momentum of the Tb$^{3+}$ ion, resulting in a total ground state angular momentum, $J = L + S = 6$, separated from the first exited multiplet, $J = 5$, by 2900~K.
The interaction with the phtalocyanine (Pc) planes further splits the energy levels in the ground multiplet.
The strong uniaxial character of this interaction leads to a ground doublet, $m_J = \pm6$, separated by about 600~K from the first exited doublet, $m_J = \pm5$. 
As we work at subkelvin temperatures, the system is confined to the two lowest states, $m_J = \pm 6$, and thus the electronic spin can be treated as an effective Ising spin 1/2. 
In a classical world the two orientations of the molecular spin would 
describe two metastable states that are separated by a large energy barrier ($\sim800$~K)~\cite{ishikawa2003lanthanide}.
However, this is not the case as non-axial interactions mix the $m_J = \pm 6$ states, with the resulting eigenstates separated in energy by an amount called tunnel splitting ($\Delta$ - Fig.~\ref{fig:zd}).
Thus, transitions between different spin orientations can occur  not only through spin lattice processes but also through quantum tunneling~\cite{gatteschi2006}.

The coupling of the molecular spin to an external magnetic field applied along the easy axis is described by the Zeeman interaction: ${\cal H}_{Z} = g_{\text{eff}}\mu_{0} \mu_{\text{B}}H_z \sigma_z$, where $g_{\text{eff}} = 18$ is the effective $g$-factor, and $\sigma_z$ is the $z$-Pauli matrix. 
 For a spin 1/2, the non-axial interactions can be modelled by an effective transverse field, $H_x = \Delta / ( g_{\text{eff}} \mu_{0}\mu_{\text{B}}) $,
so the total Hamiltonian is: 
${\cal H} = g_{\text{eff}} \mu_{0}\mu_{\text{B}}(H_z \sigma_z + H_x \sigma_x)$. 

The time evolution of the magnetic moment under a changing magnetic field is given by the following master equation for the density matrix ($\rho$):
\begin{equation}
\label{eq:do}
\frac{d\rho}{dt}= \frac{i}{\hbar}[\rho, \cal H]
\end{equation}
Thus, the spin reversal probability, for the case when the magnetic field is swept at a constant rate, is given by the Landau-Zener expression~\cite{landau1932,zener1932,stueckelberg1932}: 
\begin{equation}
\label{eq:lzs}
 P_{\text{LZ}} = 1-\exp \left(-\frac{\pi\Delta^2}{2\hbar\mu_0\mu_B|\delta m|\alpha}\right)
\end{equation}
where $\alpha$ is the field sweep rate and $\delta m$ is the change of the angular momentum upon tunneling.  

At this point, two important observations need to be made.
First, the Tb$^{3+}$ ion has a nucleus with a non-zero spin, $I=3/2$.
The strong interaction between the electronic shell and $^{159}$Tb nucleus is described by the hyperfine term, ($A_{\text{hyp}}\mathbf{I} \cdot \mathbf{J}$), and the quadrupolar term, ($P_{\text{quad}}I_z^2$), added to the total Hamiltonian.
The resulting hyperfine structure is shown in Fig.~\ref{fig:zd}b.
Thus, tunneling transitions within the doublet $m_J = \pm 6$ happen between the hyperfine states $\ket{+6,m_I}$ and $\ket{-6,m_I'}$, with $m_I$ and $m_I'$ taking values between $-3/2$ and $3/2$.
However, in a first approximation, we consider that tunneling transition events at different crossings are independent of each other, that is, tunneling dynamics at a specific anticrossing affects only the populations of the two levels that form it.

Second, when describing the physics of crystals of SIMMs, Eqs.~\ref{eq:do} and ~\ref{eq:lzs} have a limited domain of applicability, adequately describing the experimental reality only when 
the characteristic time of the experiment
(the time the system is driven through resonance) is 
considerably smaller than the characteristic time of the environmental perturbation
(\textit{e.g.} dephasing time).
The deviations from the Landau-Zener formalism are thus due to 
both elastic (dephasing) and inelastic (relaxation and excitation) processes.
The problem was recently analyzed using measurements performed on a single TbPc$_2$ molecule in a spin transistor geometry~\cite{troiani2017landau} and it was concluded that, in the limit of small probing currents, the dephasing processes dominates the system's dynamics. 
The observed behaviour was successfully modeled by a phenomenological Lindblad operator: 
\begin{equation}
\label{eq:lo}
L(t) = \frac{1}{\tau_{\text{av}}} \int_{t-\tau_{\text{av}}}^{t+\tau_{\text{av}}} {\frac{\eta}{2 \sqrt{\tau_{\text{d}}}}}[\Ket{\epsilon_1(t)}\Bra{\epsilon_1(t)}-\Ket{\epsilon_2(t)}\Bra{\epsilon_2(t)}] d\tau
\end{equation}
where $\Ket{\epsilon_{1,2}(t)}$ are the time-dependent eigenstates of $\cal H$, 
$\eta = \Bra{\epsilon_1(t)}\sigma_z\Ket{\epsilon_1(t)}$,
$\tau_{\text{d}}$ is the characteristic dephasing time representing 
the efficiency of the dephasing process. 
The time constant, $\tau_{\text{av}}$, was introduced as an interpolation parameter between the two limiting cases in which the environment affects the superposition of the diabatic or the adiabatic states~\cite{novelli2015,troiani2017landau}.
The dephasing process acts through
the term $(2L \rho L^\dagger - L^\dagger L \rho - \rho L^\dagger L)$ added to 
the right hand side of Eq.~\ref{eq:do}. 
It should be noted that, previous treatments of the dissipative Landau-Zener problem are known to theory~\cite{saito2007dissipative}, however the main advantage of the above presented formalism lies in the ability to study the decoherence process without requiring the detailed knowledge of the coupling between the molecular spin and the environmental degrees of freedom.

\textit{Results.---} 
With the above considerations in mind, we can start to analyze the 
 magnetic hysteresis loop of a crystal containing TbPc$_2$ SIMMs diluted in a diamagnetic, isostructural matrix formed by YPc$_2$ molecules, with [TbPc$_2$]/[YPc$_2$] ratio of 1\%, measured by microSQUID technique~\cite{wernsdorfer2009} (Fig.~\ref{fig:zd}a). 
Starting with a saturated sample, at a large negative magnetic field, and sweeping through the \textit{zero field} hyperfine resonances (the level anticrossings in Fig.~\ref{fig:zd}b), approximatively 85~\% of the TbPc$_2$ SIMMs undergo quantum tunneling transitions, seen as sharp steps in the magnetization curve. The remaining SIMMs reverse their magnetic moment at larger magnetic fields by a direct relaxation process~\cite{ganzhorn2014}.
Quantum tunneling transitions take place between mixed states of nuclear and electronic origin, thus both spin projections can change.
There is no relaxation step at zero field as the $I = 3/2$ nuclear spin couples to the $J=6$ electronic spin resulting in a half total integer spin and, according to Kramer's theorem, the states $\ket{+6,m_I}$ and $\ket{-6,-m_I}$, which cross at zero field, are degenerate~\cite{wernsdorfer2002s}.
The rest of the transitions can be labeled by the change in the nuclear magnetic moment ($\Delta m_I = 0,1 \text{ and }2$, shown in Fig.~\ref{fig:zd}b as a square, circle and triangle, respectively). 

\begin{figure}
    \centering
    \begin{minipage}{0.48\textwidth}
        \includegraphics[width=.99\textwidth]{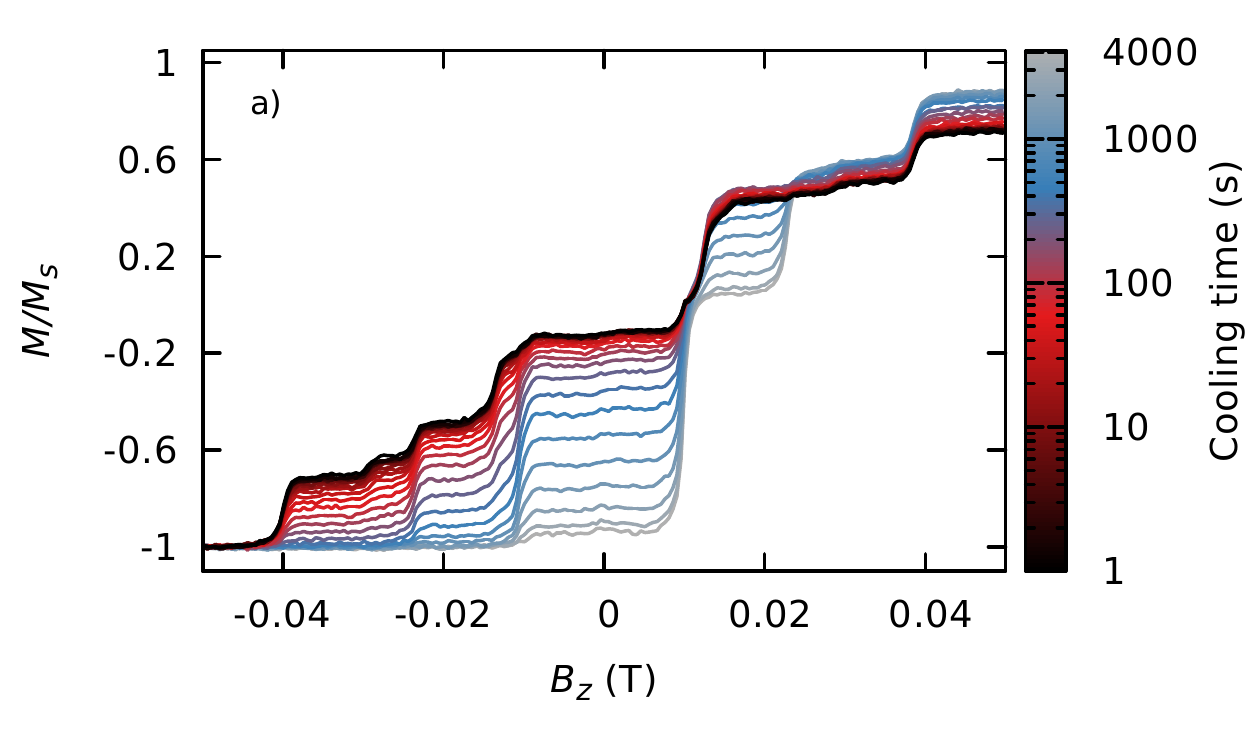}
    \end{minipage}
    \vfill
    \begin{minipage}{0.48\textwidth}
        \includegraphics[width=.99\textwidth]{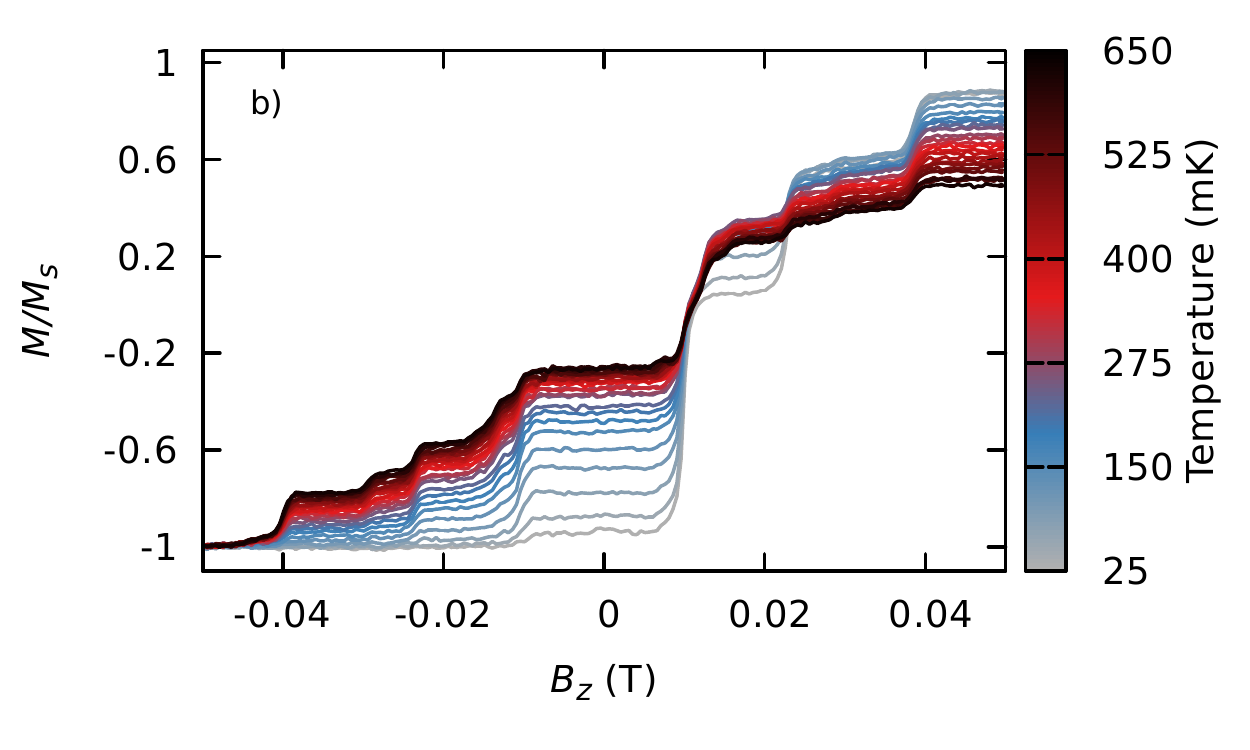}
    \end{minipage}
    \caption{a) Magnetization curves when the cryostat temperature is 25~mK, as a function of the time the sample is kept saturated in an external field of 1.3~T (cooling time). As the nuclear spins relax towards the ground state the steps that correspond to crossing involving an excited level gradually diminish and eventually disappear. b) Equilibrium magnetization curves at different cryostat temperatures obtained after the sample was kept polarized in 1.3~T for 4000s.}
	\label{fig:therm}
	\vspace{-15pt}
\end{figure}

The step heights depend both on the initial population of the hyperfine levels and on the tunneling probability.
From the Zeeman diagram in Fig.~\ref{fig:zd} one can see that the strong hyperfine and quadrupolar interaction result in exited states separated from the ground state by 122~mK, 284~mK and 476~mK. 
These spacings are larger than the lowest temperature of about 25~mK reached with our dilution cryostat. 
This suggests that any initial distribution of the nuclear spin population should evolve towards the equilibrium Boltzmann distribution. 

To show this experimentally, we first sweep the field back and fourth in the region where tunneling transitions are observed ($\sim[-0.05, 0.05]$~T) so that the nuclear spins are heated up. 
We then saturate the sample in a field of $-1.3$~T and wait for the nuclear spins to thermalize.  
Finally, we test the nuclear distribution by sweeping the zero field transitions once more. 
As the waiting time increases one observes that the steps that correspond to transitions starting from an excited state gradually diminish and then disappear (Fig.~\ref{fig:therm}a). 
At the lowest temperature of the cryostat of $\sim25$~mK a waiting time larger than $5000 \text{ s}$ is needed in order for the system to be completely thermalized, while for temperatures above 100~mK, a waiting time of about 2000~s is sufficient (Fig.~\ref{fig:therm}).
Thus, if we thermalize the sample, we know the distribution of the spins on the hyperfine states and we can use the height of the relaxation steps to obtain the tunneling transition probabilities at different crossings.

\begin{figure}
    \centering
    \begin{minipage}{0.48\textwidth}
        \includegraphics[width=.99\textwidth]{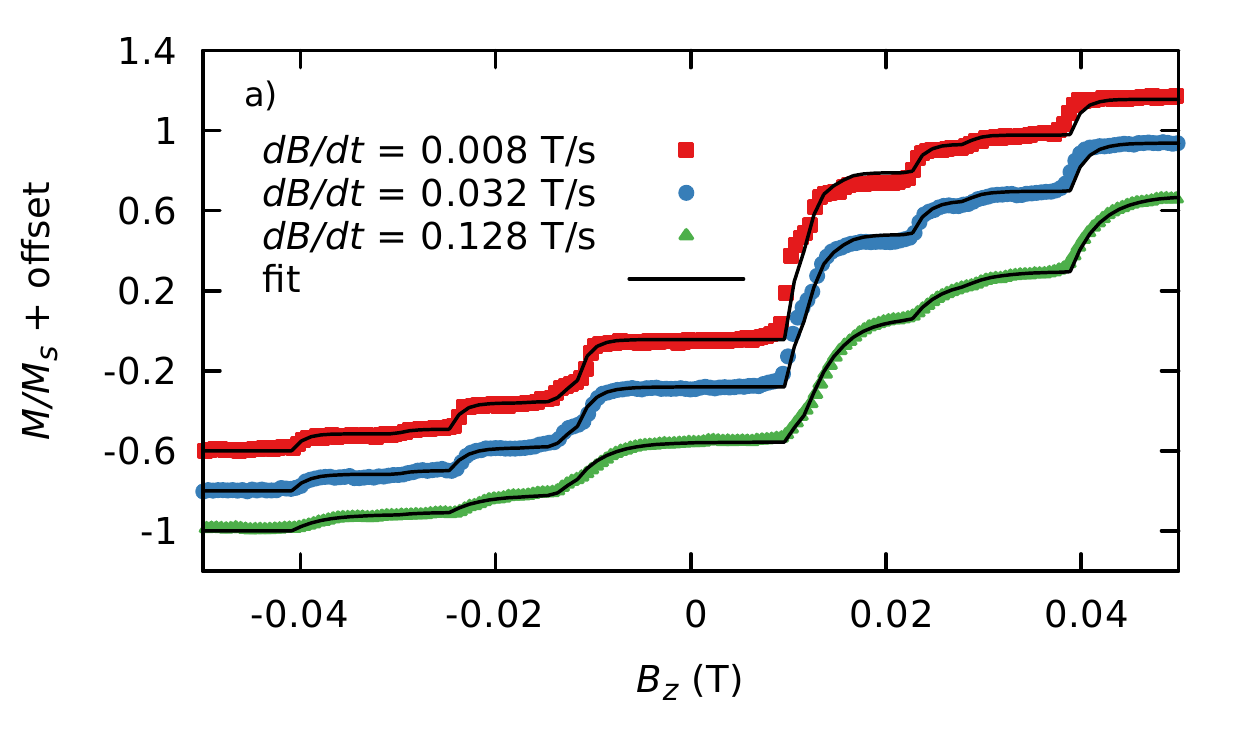}
    \end{minipage}
    \vfill
    \begin{minipage}{0.48\textwidth}
        \includegraphics[width=.99\textwidth]{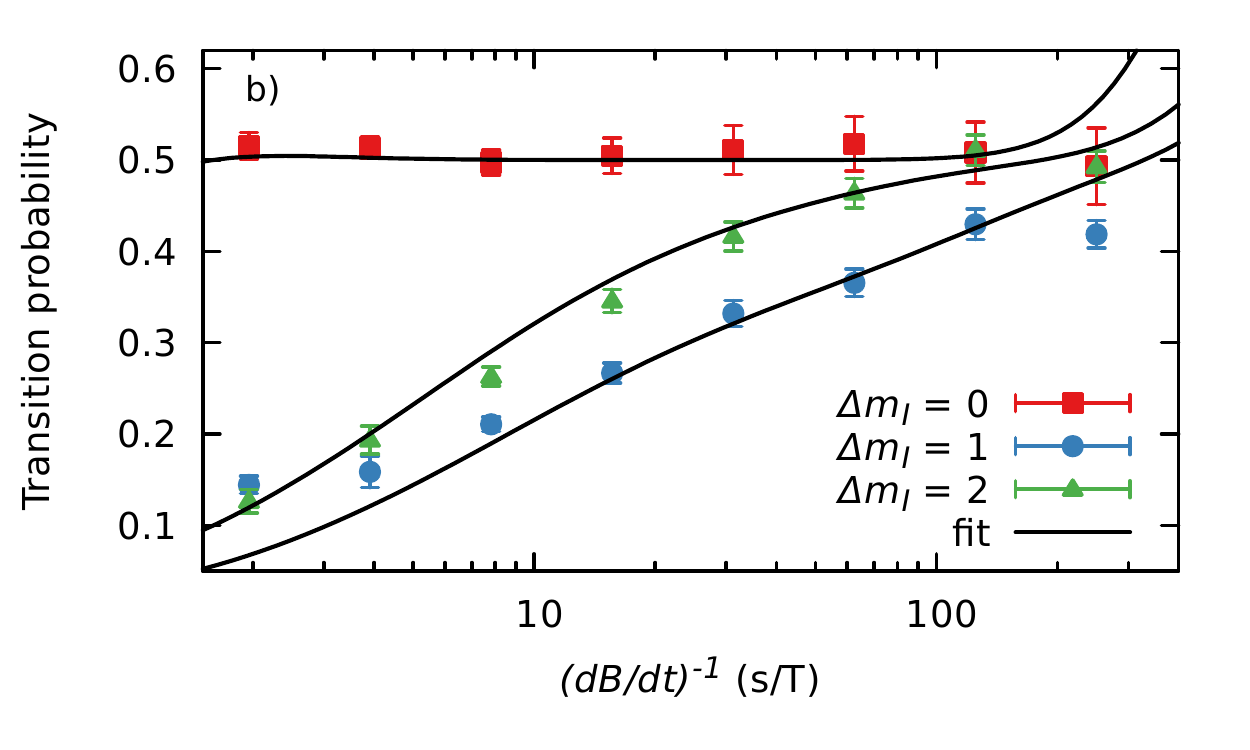}
    \end{minipage}
    \caption{a) Magnetization curves for different sweeping rates, at 200~mK, fitted by theoretical curves (black lines). The fitting procedure leads to the spin flip probability corresponding to the 3 types of anticrossings ($\Delta m_I = 0,1 \text{ and }2$). The curves were shifted vertically for a better visualization. b) Spin flip probabilities fitted to the dephasing model outlined in the text, leading to tunneling times of 4.78 and 3.47 $\upmu$s for $\Delta m_I = 1,\text{ } 2$,  respectively and a dephasing time of 0.33 $\upmu$s.
	}
	\label{fig:m_fit}
    \vspace{-15pt}
\end{figure}

The usual approach to investigate tunneling dynamics in molecular magnets is to measure the sweeping rate dependence of the relaxation steps. 
Figure~\ref{fig:m_fit}a shows the TbPc$_2$'s magnetization curves obtained after a waiting time of 2000~s at the cryostat temperature of 200~mK for 3 distinct sweeping rates. 
% magnetization fits and fitted probabilities 
To extract the transition probabilities and compare it to the Landau Zener formalism we 
fit the magnetization curves. 
The model assumes an initial Boltzmann distribution and 
that relaxation processes around zero field are entirely described by 
tunneling spin flip transitions that occur with 
probability  $P_i$ (corresponding to $\Delta m_I = \pm i$). 
The 3 tunneling parameters, $P_i$, are the only fit parameters used to model the magnetization curves (Fig.~\ref{fig:m_fit}a). 
The relation between the magnetization steps and the tunneling probabilities is: 
$ P_i = \Delta M/(2M_{\text{in}})$ where $M_{\text{in}}$ is the initial magnetization corresponding to the two levels that form the anticrossing and $\Delta M$ is the step height.
Thus, sweeping over a level anticrossing formed by two states, $\Ket{+6, m_I}$ and $\Ket{-6, m_I'}$, with initial relative populations $n_i$ and $n_j$, modifies the population of the hyperfine levels in  accord with the following relation: $n_{i,j}' = n_{i,j} \mp P_{|m_I - m_I'|} (n_i - n_j)$.

There are 2 additional factors that come into play when modeling the measured magnetization curves. 
First, the steps are broadened due to dipolar interactions in the system
(from experimental curves this is taken to be around $\sqrt{\Delta H_{\text{dip}}^2} = 1$~mT).
Note that, if the dipolar field is quasistatic during the relaxation process of the individual molecular spins (as it is the case for diluted samples and small relaxation steps), the effect of the dipolar coupling is just to shift the time origin of the Landau-Zener process, with no effect on the dephasing process.
Then, the steps are artificially broadened due to the time constant of the feedback loop of the measurement process~\cite{wernsdorfer2009}. 
Taking into account the above two factors results in an almost perfect fit of the experimental curves, as seen in Fig.~\ref{fig:m_fit}a. 

The determined transition probabilities are shown in Fig.~\ref{fig:m_fit}b.
It is important to notice that at very slow sweeping rates ($dB/dt \lessapprox 5$~mT), the model for the magnetization curves starts to fail because the characteristic time of the experiment approaches the electronic and nuclear spin lattice relaxation times. Thus, we work in the sweeping rate range in which the tunneling dynamics dominates the direct relaxation process.

In order to characterize the dephasing process, we solve numerically the above presented phenomenological master equation and fit the spin flip transition probabilities using a nonlinear least-square algorithm with three time constants: $\tau_{\Delta}$, $\tau_{\text{d}}$ and $\tau_{\text{av}}$. 
As seen in Ref.~\onlinecite{troiani2017landau}, if we are far from  the coherent Landau-Zener dynamics, then $\tau_{\text{d}}/\tau_{\Delta}$ and $\tau_{\text{av}}/\tau_{\Delta}$ parameters are uniquely defined by the shape of the $P(dB/dt)$ characteristic, while a variation of the tunneling time, $\tau_{\Delta}$, results in a horizontal shift along the sweeping rate axis.
Additionally, to reduce the number of fit parameters, we make the requirement for $\tau_{\text{d}}$ and $\tau_{\text{av}}$ to be the same for $\Delta m_I = \pm 1 \text{ and } \pm 2$ transitions.

The transitions that conserve the nuclear spin ($\Delta m_I = 0$) are independent on 
the sweeping rate (in the range tested experimentally).
The plateau at $P_0=0.5$ is characteristic of a dephasing process that comes from
a strong interaction between the system and its environment, so that 
the Lindblad operator is mostly constructed from
the diabatic states~\cite{troiani2017landau}.
The $\Delta m_I = 1\text{ and }2$ transitions are well fitted by using
tunneling times, ($\tau_{\Delta} \equiv \hbar/\Delta$), $\tau_{\Delta}=4.78\text{ and }3.47$ $\upmu$s,  respectively, and
with the same dephasing ($\tau_{\text{d}} = 0.33$ $\upmu$s or alternatively a decoherence rate $\gamma_{\text{d}} = 1/\tau_{\text{d}} \approx 3$~MHz) and averaging time ($\tau_{\text{av}} = 93.7$ $\upmu$s).
The sweeping rate range for which the plateau of $P_0=0.5$ is observed for $\Delta m_I = 0$ means that the tunnel splitting for this anticrossing is at least 10 times larger than the other 2 transitions
(the fit curve is not uniquely defined for $\Delta m_I = 0$ transition).

The study in Ref.~\onlinecite{troiani2017landau} and the present work share the same molecular complex but which is placed in very different environments.
Thus, it is worthwhile to compare the measured low temperature dynamics. 
For the TbPc$_2$ molecule in a spin transistor geometry, tunneling events are observed only at the crossings that conserve the nuclear spin while for TbPc$_2$ in a single crystal environments, all the transitions except the ones at zero field are evidenced experimentally~\cite{ishikawa2005quantum, taran2019therole}.
This clearly shows that the molecules in the two samples are acted upon by different transverse terms.
Also, the measured $P(dB/dt)$ characteristics differ significantly between the two experiments proving that the dynamics of a molecular spin driven though an avoided level crossing is strongly dependent on the coupling to its environment. 
In the former study, the conduction electrons that tunnel through the ligand quantum dot are expected to play the dominant role in the decoherence process, while for a molecular crystal at very low temperatures, the incoherent dynamics is caused mainly by the surrounding spin bath comprised of nuclear and other molecular spins.
Establishing the connection between the phenomenological model that uses Lindblad operators and a microscopic description that includes explicitly the environmental degrees of freedom is an important outlook of the present study.

\textit{Conclusions.---}
Even though there still remain a number of open questions regarding 
decoherence in mesoscopic systems~\cite{stamp2012environmental},
both theoretical and experimental progress has been reported~\cite{morello2006pairwise,takahashi2011decoherence}.
This article represents another step towards understanding
the complex dynamics of an ensemble of interacting quantum systems, so that
we can get closer to functional devices that make use of their properties.
We show that the incoherent Landau-Zener dynamics can be used to infer relevant information regarding the quantum dynamics in crystals of molecular magnets. The combination of low temperature magnetometry with appropriate theoretical tools has the potential to complement the resonant techniques used so far. 

\section*{Acknowledgement}
We acknowledge the Alexander von Humboldt Foundation and the ERC advanced grant MoQuOS No. 741276.
%Bibliography
\bibliographystyle{apsrev4-1}
\bibliography{bib}

\end{document}